# An *HST*/WFC3 view of stellar populations on the horizontal branch of NGC 2419

M. Di Criscienzo,[1]★ M. Tailo,[1] A. P. Milone,[2] F. D'Antona,[1] P. Ventura,[1] A. Dotter[2] and E. Brocato[1]

[1]*INAF, Osservatorio Astronomico di Roma, Via Frascati 33, Monteporzio Catone (Roma), Italy*
[2]*Research School of Astronomy & Astrophysics, Australian National University, Mt Stromlo Observatory, Weston, ACT 2611, Australia*

## ABSTRACT

We use images acquired with the *Hubble Space Telescope* Wide Field Camera 3 and new models to probe the horizontal branch (HB) population of the Galactic globular cluster (GC) NGC 2419. A detailed analysis of the composite HB highlights three populations: (1) the blue luminous HB, hosting standard helium stars ($Y = 0.25$) with a very small spread of mass; (2) a small population of stars with intermediate helium content ($0.26 < Y \lesssim 0.29$); and (3) the well-populated extreme HB. We can fit the last group with models having high helium abundance ($Y \sim 0.36$), half of which (the hottest part, 'blue hook' stars) are identified as possible 'late flash mixed stars'. The initial helium abundance of this extreme population is in nice agreement with the predicted helium abundance in the ejecta of massive asymptotic giant branch (AGB) stars of the same metallicity as NGC 2419. This result further supports the hypothesis that second-generation stars in GCs formed from the ashes of intermediate-mass AGB stars. We find that the distribution in magnitude of the blue hook stars is larger than that predicted by theoretical models. We discuss the possible uncertainties in the magnitude scales and different attempts to model this group of stars. Finally, we suggest that consistency can be better achieved if we assume core masses larger than predicted by our models. This may be possible if the progenitors were fast rotators on the main sequence. If further study confirms this interpretation, a fast initial rotation would be a strong signature of the peculiarity of extreme second-generation stars in GCs.

**Key words:** stars: horizontal branch – globular clusters: general – globular clusters: individual: NGC 2419.

## 1 INTRODUCTION

NGC 2419 is among the most distant (∼87.5 kpc) and luminous ($M_V = -9.5$) globular clusters (GCs) of the Galaxy's outer halo. Its half-light radius ($r_h \sim 24$ pc; Ibata et al. 2011) is by far larger than that of other GCs of the same luminosity and more similar to that of a dwarf galaxy. For these reasons, the attention given to NGC 2419 has grown considerably in the last few years with the final scope to analyse the connection between GCs and the primordial building blocks in the hierarchical merging paradigm of galaxy formation.

Ripepi et al. (2007) and Di Criscienzo et al. (2011a) used ground-based photometric data to highlight the complex structure of the horizontal branch (HB) of NGC 2419. Thanks to space data from the *Hubble Space Telescope* (*HST*), di Criscienzo et al. (2011b, hereafter DC11) could attempt a detailed analysis of the HB morphology, finding that the HB is composed mainly of two separate groups of stars: (1) the luminous blue HB stars that extend by evolution into the RR Lyrae and red HB region and (2) the extreme HB (EHB), populated by hotter and fainter stars. DC11 concluded that the only viable explanation for the latter group of stars is that they belong to a helium-rich ($Y \sim 0.42$), second generation (SG) of stars, formed within ∼200 Myr since the formation of the main component. The colour dispersion of the red giant branch (RGB) was also consistent with this hypothesis, as recently confirmed by Beccari et al. (2013) using a large number of deep multi-band images from the LBC camera on the Large Binocular Telescope.

On the spectroscopic side, the recent works of Mucciarelli et al. (2012) and Cohen & Kirby (2012) confirm that no intrinsic spread of iron is present in NGC 2419 ([Fe/H] = $-2.09 \pm 0.02$ dex), as for a typical GC. On the other hand, both the Mg and K abundances display large variations, strongly anticorrelated with each other. The K variation has never have been observed before, neither in GCs nor in dwarf galaxies, stressing the peculiarity of NGC 2419.

★ E-mail: dicrisci@gmail.com

In the framework of the model in which the multiple GC generations are due to self-enrichment from massive asymptotic giant branch (AGB) and super-AGB stars (Ventura et al. 2001; D'Ercole et al. 2008), Ventura et al. (2012) made a significant step forward in explaining the chemically peculiar features observed in this cluster. They showed that the high potassium abundance found in the Mg-poor stars can be achieved via hot bottom burning (HBB) during the evolution of massive ($M \simeq 6$ $M_\odot$) AGB stars: potassium is formed by p-captures on argon nuclei, either if the relevant cross-section(s) are larger than those listed in the literature or if the HBB temperature is slightly larger than that predicted by the models. Carretta et al. (2013), comparing the abundances of Mg and K with those of other GCs and field stars, conclude that at present NGC 2419 seems unique among GCs. This may be due both to the low metallicity of this cluster (leading to larger HBB temperatures in the AGB phase) and to the formation of SG stars from the masses which experience the strongest HBB. In this case, it would be very difficult to explain the huge helium abundance ($Y = 0.42$) predicted by DC11 for the SG stars, as the maximum helium abundance achieved by AGB models is $Y \simeq 0.35$–$0.37$.[1] In this work, we look in detail at this question by taking advantage of new isochrones and HB models as well as new ultraviolet (UV) photometry from the *HST* Wide Field Camera 3 (WFC3): well suited to the analysis of the core He-burning stars with effective temperatures hotter than $T_{eff} \simeq 25\,000$ K (i.e. the EHB stars). After core helium burning, these stars, owing to the small masses of their envelopes, evolve into AGB manqué stars (or post-early AGB stars) and do not return to the AGB (e.g. Dorman, Rood & O'Connell 1993). EHB stars are found in many moderately metal poor GCs such as NGC 6752 (Heber et al. 1986), NGC5986 and M80 (Moni Bidin et al. 2009). In addition, some massive GCs (e.g. $\omega$ Cen, D'Cruz et al. 1996 and NGC 2808, Brown et al. 2001) show a peculiar class of objects that form a blue hook (BHk) at the hottest end of the EHB (D'Cruz et al. 1996, 2000; Brown et al. 2001). Following the pioneering work by Castellani & Castellani (1993), D'Cruz et al. (1996) proposed that such stars may have suffered a delayed helium-core flash. In these 'late flashers' (LFs), the mass of the envelope is reduced to a level that H-burning cannot energetically sustain the star: the flash starts away from the RGB tip when contraction of the external layers has already begun. The different cases of the late flash events have been thoroughly described in the literature (see Brown et al. 2001). In some LFs, ignition occurs while the star is already on the helium white dwarf (WD) cooling sequence: in this case, the event is more dramatic and initiates a deep mixing event that leaves a helium burning core surrounded by a residual envelope, enriched in helium (mass fraction >90 per cent) and carbon ($\sim$2–3 per cent) synthesized during the helium flash. The core mass of the 'late flash mixed stars' (LFMs) is smaller than the other LFs. Among EHB stars, LFMs are several thousand kelvin hotter, owing to the peculiar compositions of their atmospheres and also to the flash mixing, which decreases the opacity below the Lyman limit (Brown et al.2001).

In $\omega$ Cen, most of the BHk objects lay along two parallel sequences, with $T_{eff}$ increasing with luminosity, and this evidence must bear a relation to the peculiar evolution of LFM stars[2] (Cassisi et al. 2009; D'Antona, Caloi & Ventura 2010); in other clusters, the observational data are less clear and so is their evolutionary interpretation. Brown et al. (2010) show that LFM models are required to explain the faint luminosity of the BHk stars in the five massive GCs NGC 2419, NGC 6273, NGC 6715, NGC 6388 and NGC 6441. According to Brown et al., neither the LF nor the LFM models can explain the full range of colour observed in such stars.

From a photometric perspective, a complete characterization of EHB stars in GCs can best be performed in UV colour-magnitude diagrams (CMDs) where, thanks to the small bolometric corrections, the atmospheric effect of enhanced helium and carbon abundances can be distinguished by means of the resulting changes in temperature (see Brown et al. 2010). With the help of new *HST* UV data, the large EHB population in NGC 2419 and detailed evolutionary HB models computed specifically for this work, we will try to shed light on this topic.

Interestingly, the EHB stars are also important for their influence on the integrated colours of stellar populations. They are one of the important sources of the UV upturn at wavelengths shorter than 2300 Å in the spectra of elliptical galaxies, which at the moment is still a controversial issue (e.g. Greggio & Renzini 1990; Dorman, O'Connell & Rood 1995; Chung, Yoon & Lee 2011; Bekki 2012). In Section 2, the new data from WFC3 are presented, in Section 3, the theoretical scenario is described and in Section 4, we show our attempts to interpret the different populations of HB stars with particular attention on the blue tail. Conclusions close the paper.

## 2 OBSERVATION AND DATA REDUCTION

Our data set consists of images collected by the *HST* WFC3 UVIS channel and is described in detail in Table 1.

Photometry and astrometry have been carried out with img2xym_WFC3 (Bellini et al. 2010), which is mostly based on img2xym_WFI (Anderson et al. 2006). Star positions have been corrected for geometrical distortion by using the solution provided by Bellini, Anderson & Bedin (2011). Photometry has been calibrated as in Bedin et al. (2005) by using the zero-points provided in the web page of the Space Telescope Science Institute.[3]

Since our analysis is focused on high-precision photometry, we have selected stars with small point spread function (PSF)-fitting errors by using the procedure described by Milone et al. (2009) and based on the quality indices that the photometric software by Bellini et al. (2010) produces. Finally, we have corrected our photometry for spatial variations of the photometric zero-point due to small uncertainties in the PSF model following the procedure described by Milone et al. (2012).

The complete CMDs reported in Fig. 1 clearly show how UV observations are the best photometric tool to study hot stellar populations that emit most of their light at short wavelengths. In fact, going from optical bands (e.g. *F438W* and *F814W*) to the UV (*F225W*), the brightest stars in the CMD are the HB stars.

Since in the following we will focus on the HB stars, in Fig. 2 we show the combination of photometric bands that, after a careful analysis, best highlights the gaps between different populations on the HB, encompassing the extremely wide ($\Delta\lambda = 753.0$ Å) *F300X*-band UV filter, centred at $\lambda = 2829.8$ Å.

The stars with $m_{F225W} - m_{F300X} < 0$ populate what we call the EHB. In the 'normal' HB population, having $m_{F225W} - m_{F300X} > 0$, a gap at $m_{F225W} \sim 21$ mag is present, as shown by the histogram in the right-hand panel of Fig. 2. This is obtained by straightening the HB sequence, which is accomplished by computing the difference in colour of each star with respect to the ridge line.

---

[1] This upper limit to the maximum helium abundance in the AGB models is independent of the details of AGB modelling – see e.g. Ventura et al. (2014).
[2] Many issues are still pending, which will be discussed in Tailo et al. (in preparation).

[3] http://www.stsci.edu/hst/wfc3/phot_zp_lbn

**Table 1.** Description of the WFC3 data sets used in this paper.

| Prop. ID | Date | RA | Dec. | Filter | Exp. time(s) |
|---|---|---|---|---|---|
| 11903 | 15-05-10 | $07^d38^m07^s.8$ | 38:52:48.0 | *F225W* | 750 |
| 11903 | 15-05-10 | $07^d38^m07^s.8$ | 38:52:48.0 | *F300X* | 467 |
| 11903 | 15-05-10 | $07^d38^m07^s.8$ | 38:52:48.0 | *F438W* | 725 |
| 11903 | 15-05-10 | $07^d38^m07^s.8$ | 38:52:48.0 | *F814W* | 650 |

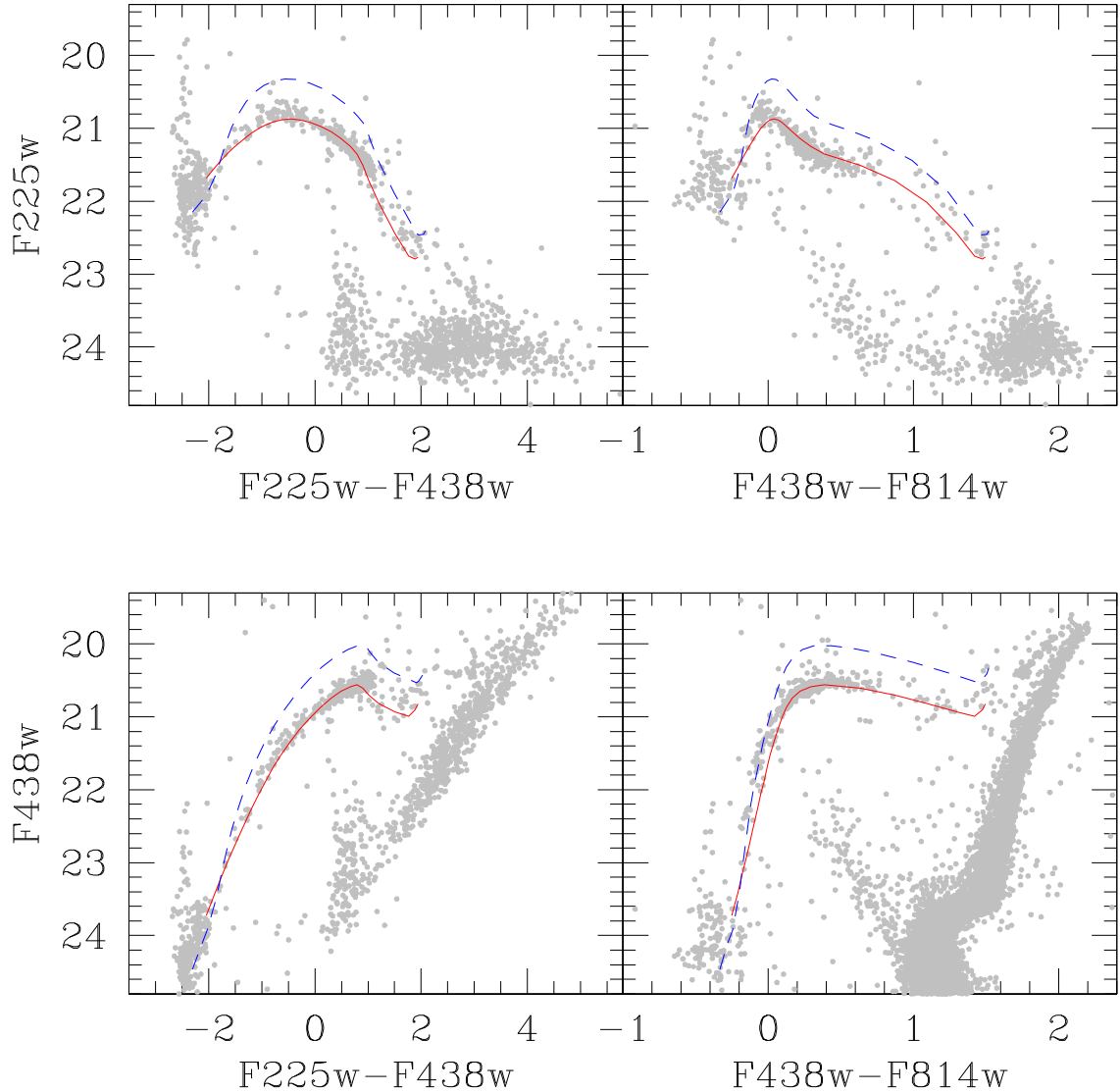

**Figure 1.** CMDs using $m_{F225W} - m_{F438W}$ (left-hand panels) and $m_{F438W} - m_{F814W}$ (right-hand panels) colours and $m_{F225W}$ (upper panels) and $m_{F438W}$ (lower panels) that show how going from optical to UV photometrical bands the HB stars became the brightest stars of the cluster. Zero-age HB for $Y = 0.25$ (red solid line) and $Y = 0.40$ (blue dashed line) are also plotted (see the text for the distance modulus and reddening used in this paper.)

According to this empirical classification, the total HB sample in this UV catalogue consists of ~450 red (RHB), ~100 intermediate (IHB) and ~350 extreme (EHB) stars. These correspond, respectively, to 50, 9 and 41 per cent of the total HB population. Among the EHB stars, we also considered ~25 stars brighter than $m_{F225W} = 21.5$, whose origin is still unknown (see Section 4 for a possible interpretation).

## 3 EVOLUTIONARY MODELS AND SIMULATIONS

The models presented in this work and used to simulate the HB have been computed with the ATON code by using the updated input physics described by Ventura et al. (2009). We assume a metallicity $Z = 0.0003$ and four different values of the initial helium ($Y = 0.25$, 0.28, 0.35, 0.40). The mixture is $\alpha$-enhanced, with [$\alpha$/Fe] = 0.4, in agreement with the recent spectroscopic results by Mucciarelli et al. (2012). We recall that DC11, based on the analysis of a very small sample of stars (Shetrone, Cote & Sargent 2001) available at that time, adopted a much smaller Z. This new choice ($Z = 0.0003$) is the same adopted by Ventura et al. (2012) to build up the yields that have been compared with the most recent chemical inventory of NGC 2419 (Cohen & Kirby 2012).

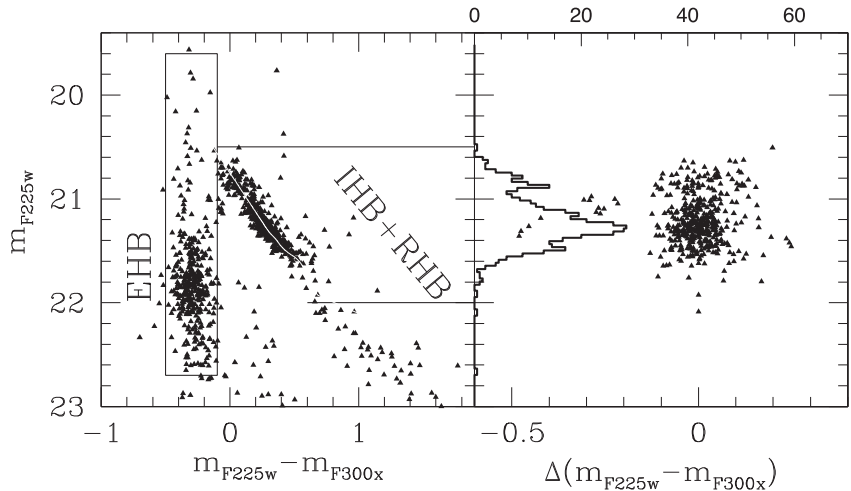

**Figure 2.** CMDs in the UV photometric bands chosen to delineate the three different parts of the HB (left-hand panel). In the right-hand panel, we show the differences computed with respect to the ridge line (white line) for the bulk of stars with $m_{F225W} - m_{F300X} > 0$. The histogram shows the presence of a gap at $m_{F225W} \sim 21$ that separates the red part (RHB) from the intermediate population (IHB).

We evolve low mass stellar models through the main sequence (MS) and RGB, to calculate the isochrones and to determine the helium-core mass at the flash ignition. The resulting core masses were taken as an input for the computation of the HB models, while the mass evolving along the RGB ($M_{RG}$), as a function of age and helium content, was used as an input for the simulations of the HB (see Table 2).

Synthetic models for HB stars were computed according to the recipes described in D'Antona & Caloi (2008). Adopting $M_{RG}$ at a given age as a function of helium (at the chosen, fixed metallicity of the cluster), the mass on the HB is $M_{HB} = M_{RG}(Y) - \Delta M$, where $\Delta M$ is the mass lost during the RG phase (or just after in the case of LF stars). Table 2 shows the difference in the evolving giant mass at an age of 12 Gyr for different helium contents. We assume that $\Delta M$ has a Gaussian dispersion $\sigma$ around an average value $\Delta M$ and that both $\Delta M$ and $\sigma$ may depend on the helium content.[4] We associate a random spread in magnitude (up to 0.015 mag) to each point, to simulate the impact of observational errors.

In order to test the late helium flash mixing scenario in the interpretation of BHk stars, we use complete evolutionary tracks in which the helium flash occurs later on the WD cooling sequence. Consistent with previous computations (Brown et al. 2001; Miller-Bertolami et al. 2008), when the mass of the H-rich envelope drops below a minimum threshold before the flash ignition in the core, the star leaves the RGB and starts contracting. During the following phase, two highly energetic ignition episodes occur: the helium flash, followed by a hydrogen flash. A convective layer develops and eventually reaches the regions processed by the two burning episodes. The result is a hydrogen-depleted star with surface helium and carbon mass fractions of about 0.96 and 0.02, respectively. Finally, after further, less energetic burning episodes, the star reaches the quiescent helium burning phase with a total and core mass smaller than its cooler helium flash counterparts.

It is worth noting that, compared to the sequences showed in Miller-Bertolami et al. (2008), almost all of our models experience a deeper mixing. This confirms previous findings, that shallow mixing events are rare in metal-poor sequences (Lanz et al. 2004).

By varying the mass-loss in our computations, we find that the mixing event occurs only in a narrow interval of mass, ranging from 0.494 to 0.499 $M_\odot$ for $Y = 0.25$, and from 0.457 to 0.464 $M_\odot$ for $Y = 0.40$. These values are reported in Table 2, where the core mass during the HB evolution and the mean helium and carbon abundances resulting from this peculiar evolution are indicated. We also show the total and core masses of the least massive model not undergoing the late helium flash mixing.

The helium burning phase of the flash mixed models occurs approximately at the same $T_{eff}$, independently of mass. However, owing to the differences in core mass, higher $Y$ models are fainter than their standard initial $Y$ counterparts. In spite of these differences, the $Y = 0.25$ ($Y = 0.40$) sequences are, on average, $\sim$13 000 ($\sim$7000) hotter than the least massive model not undergoing the late helium flash mixing (see Fig. 3).

We include the LFM evolution in our synthetic simulations of the HB of NGC 2419, by assuming that, for each $Y$, there is a boundary mass separating the (larger) masses ending on the zero-age HB (ZAHB, without mixing) and the (smaller) masses undergoing the mixing (see third and fifth columns in Table 2); below the less massive LFM, the star is assumed to evolve into the helium WD sequence, and is excluded from the number of EHB stars.

### 3.1 Assembling together all ingredients in the CMD

All the magnitudes were transformed to the VEGAMAG photometric system using the procedure described by Holtzman et al. (1995), with zero-points listed in the *HST* Data Handbook (table 28.1).

To find out distance modulus and reddening, we fit the lower envelope of the red side of the HB with the ZAHB corresponding to $Y = 0.25$: we find $(m - M)_o = 19.63$ mag and $E(B - V) = 0.08$. By assuming $A_V = 3.1 \times E(B - V)$, we have determined the

---

[4] The most straightforward hypothesis is that mass-loss and dispersion in mass-loss *do not* depend on the helium content (D'Antona et al. 2002; D'Antona & Caloi 2004). A collection of recent results shows that a mass-loss larger by a few thousandths of solar mass may be required for SG stars (see e.g. Salaris, Cassisi & Pietrinferni 2008; D'Antona et al. 2013).

**Table 2.** Properties of HB models calculated for this work and available for HB population synthesis. In particular initial helium abundance (*Y*), corresponding evolving mass on RGB at 12 Gyr ($M_{12\,\mathrm{Gyr}}$), minimum mass on HB ($M_{\mathrm{min, HB}}$) and its core mass ($M_{\mathrm{core, He}}$) are reported. For $Y = 0.25$ and 0.40, we give the total and core masses range for which we obtain the mixing events (see the text) together with the final chemistry (mean helium and carbon mass content) of the LF mixed stars.

| Standard HB models | | | | LF mixed models | | | |
|---|---|---|---|---|---|---|---|
| Y | $M_{12\,\mathrm{Gyr}}$ | $M_{\mathrm{min, HB}}/\mathrm{M}_\odot$ | $M_{\mathrm{core, He}}/\mathrm{M}_\odot$ | $M_{\mathrm{LFM}}/\mathrm{M}_\odot$ | $M_{\mathrm{core, LFM}}/\mathrm{M}_\odot$ | $Y_{\mathrm{mix}}$ | $C_{\mathrm{mix}}$ |
| 0.25 | 0.8111 | 0.50 | 0.497 | 0.494–0.499 | 0.484–0.494 | 0.965 | 0.025 |
| 0.28 | 0.7701 | 0.50 | 0.489 | – | – | – | – |
| 0.35 | 0.6785 | 0.48 | 0.471 | – | – | – | – |
| 0.40 | 0.6168 | 0.46 | 0.456 | 0.457–0.464 | 0.445–0.454 | 0.963 | 0.024 |

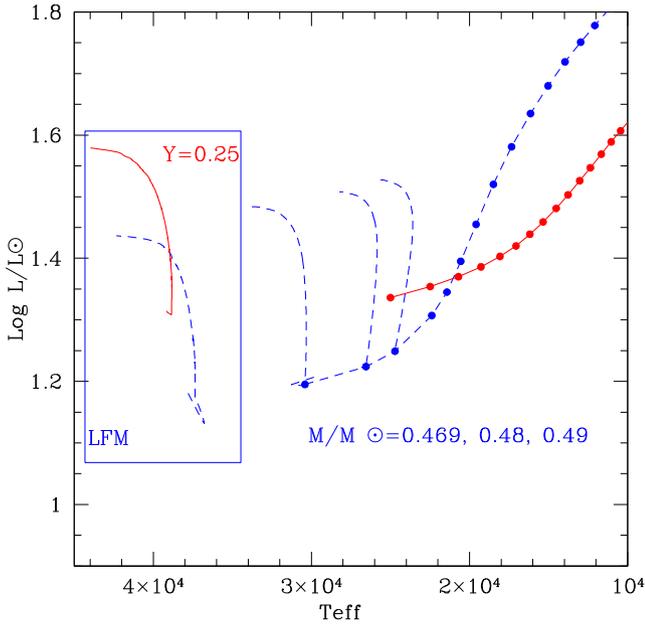

**Figure 3.** Evolution of HB for the late helium flashers mixed with $Y = 0.25$ (solid line in the rectangular box) and $Y = 0.40$ (dashed line in the rectangular box) compared with the 'normal' ZAHB. In the case of $Y = 0.40$, the HB evolution of less massive models is also shown.

## 4 CLUES ON THE NATURE OF DIFFERENT GROUPS OF HB STARS

We use synthetic HB populations, based on the models described above, to better understand the main features of the HB of NGC 2419 and to give an interpretation of the division into at least three parts (RHB, IHB and EHB) as pointed out in Section 2.

First, we confirm the result of DC11 about the RHB: it is composed of a single population, with a very narrow mass interval ($\delta M = 0.005 \,\mathrm{M}_\odot$). We fix the cluster age at 12 Gyr (which reproduces the location of SGB+RGB stars at the assumed distance and reddening[6]), at which the mass evolving along the RGB is 0.811 $\mathrm{M}_\odot$. An average mass-loss of $\Delta M = 0.14 \,\mathrm{M}_\odot$ along the RGB leads to an average mass on the HB of 0.67 $\mathrm{M}_\odot$. As a consequence of the higher metallicity adopted in the present investigation, the mass-loss is higher than in DC11 ($\sim 0.073 \,\mathrm{M}_\odot$).

As shown in Section 2.1, the new data – particularly the gap at $m_{F225W} = 21$ mag – allow us to isolate an IHB population. The comparison with the ZAHBs computed with different helium content clearly suggests that at least 9 per cent of the luminous HB stars have a helium abundance higher than the standard value. We are able to reproduce the observed distribution of stars by assuming a spread in the helium abundance of $\Delta Y < 0.03$ and an average mass-loss slightly larger ($\Delta M = 0.17 \,\mathrm{M}_\odot$) than in the brighter part. In Fig. 4, we show the promising comparison of this simulation with the data, where both the observed RHB and the IHB are reproduced.

In contrast, the small enhancement in *Y*, sufficient to match the IHB stars, is not enough to reproduce the location of the EHB stars. To this aim, DC11 adopted $Y = 0.42$ (and a larger mass-loss on the RGB). The higher metallicity adopted in this work reduces the *Y* necessary to fit the data, compatible with the results of AGB model predictions. Adopting $Y = 0.36$ [as from the models by Ventura et al. (2012), which have the strongest HBB], only a slightly larger mass-loss ($\Delta M \sim 0.190 \,\mathrm{M}_\odot$) is necessary to reproduce the luminosity of the bulk of the EHB stars in $m_{F225W} - m_{F300X}$ colours. This result makes the AGB – super-AGB – models fully compatible not only with the spectroscopic signatures, but also with the photometric data of this cluster.

Using the quoted value for the mass-loss (see Sim 1 in Table 3), only the faintest stars (with $m_{F225W} > 22.4$) are interpreted as LFMs. However, the same simulation does not fit the data when a larger colour baseline is adopted, as shown in detail in the right-hand panels of Fig. 5, where we use the $m_{F225W} - m_{F438W}$ colour. In particular, a group of brighter stars having the same $m_{F225W} - m_{F300X}$ filter-specific extinction as a function of effective temperature. This is achieved by integrating synthetic spectra combined with the Cardelli, Clayton & Mathis (1989) extinction curve over the throughput of each filter.

In the case of LFM models, for consistency, we should consider atmospheres calculated with a mixture enhanced in carbon and helium. By convolving the spectra kindly provided by Tom Brown with the transmission curves of WFC3, we found that the carbon enhancement, at the considered temperatures, makes the $m_{F225W} - m_{F300X}$ colour about 0.04 mag redder, and $m_{F225W} \sim 0.06$ mag fainter. Unfortunately, the available spectra are limited to wavelengths below 4000 Å, which prevented us from calculating the correction factor in the $m_{F225W} - m_{F438W}$ colour that will be used in the next section; we simply assumed that it cannot exceed 0.06 mag.[5] In our models, we have taken into account this systematic effect by shifting the LFM stars by 0.06 mag to the red.

---

[5] We consider this value as an upper limit, since the surface carbon abundance has been increased to 3 per cent in the flash mixed models provided by Brown, while our models attain 2 per cent abundance (see Table 2).

[6] Different choices for the age are possible, but the differential results are very similar, once the mass-loss is adjusted to locate the first-generation HB stars at their observed location.

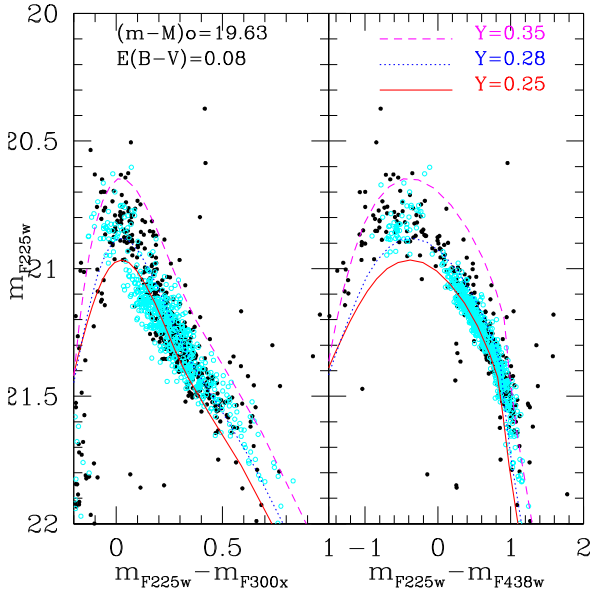

**Figure 4.** We show the CMD of the RHB+IHB in two WFC3 colours, together with the ZAHBs of $Y = 0.25$, 0.28 and 0.35. We show in cyan colour the simulation described in the text.

**Table 3.** Details about the mass-loss (centre of the Gaussian distribution $\Delta M$ and $\delta$) and the helium content of EHB+LFM simulations discussed in the text. The number of LFM predicted by each simulation is also shown.

| | EHB | | | |
|---|---|---|---|---|
| Name | $Y$ | $\Delta M$ | $\delta$ | $N_{\mathrm{LFM}}$ |
| Sim 1 | 0.36 | 0.190 | 0.003 | 20 |
| Sim 2 | 0.36 | 0.195 | 0.003 | 130 |
| Sim 3 | 0.25+0.36 | 0.323+0.185 | 0.003 | 130 (all with $Y = 0.25$) |

colour of the simulated LFM stars shows up. In this CMD, a gap at $m_{F225W} - m_{F438W} = -2.25$ mag, dividing the bluest part of the HB into two equally populated groups, is clearly visible.

In the upper-right panel of Fig. 5, we show the ZAHBs and the LFM tracks for different helium contents. In the $m_{F225W} - m_{F438W}$ colour, it seems more plausible that *half* (the bluest group) of the EHB stars are LFM stars. We built a new simulation (Sim 2) with a slightly larger mass-loss (0.195 M$_\odot$) than in the previous case, which allows us to obtain the same numbers of BHk and EHB; however, the comparison with the observations shows that the bluest part of the HB is too faint, when the same helium used to describe the red side of EHB is adopted.

We obtain a better fit of the data if we assume that LFMs are first-generation stars, with helium $Y = 0.25$[7] taking the hypothesis that stars in the red group have a large initial helium content ($Y = 0.36$; see Sim 3 in Table 3). The problem with this interpretation can be easily understood by looking at the values of the evolving masses of Table 2: for $Y = 0.25$ at 12 Gyr, it is 0.811 M$_\odot$. The average mass-loss $\delta M = 0.14$ M$_\odot$ brings the post-RGB evolution into the luminous part of the HB, which is extremely well populated. Therefore, we should add the hypothesis that a non-negligible fraction of the same stars (~15 per cent) suffer a well-defined mass-loss (0.32 M$_\odot$), so that they all pass through an LFM episode and join the location of the high-helium LFM population. How plausible is the idea that violent encounters among stars of the first generation (see e.g. Adams, Davies & Sills 2004; Pasquato et al. 2014) bring the helium burning stars to the LFM phase as observed in NGC 2419? In our opinion, the continuity with the EHB populations in the CMD suggests a common history for both groups of stars. As already demonstrated in DC11 (and confirmed in the upper-right panel of Fig. 5 of this work), an enhanced helium abundance is mandatory to reproduce the magnitude and the colour of EHB stars; for this reason, we propose in the next section a different scenario.

---

[7] The same hypothesis was put forward by Cassisi et al. (2009) to interpret the wide luminosity span of the BHk stars in $\omega$ Cen.

### 4.1 A possible spread in the core masses of BHk stars due to rotation

Several observations of GC stars suggest that the mass-loss may be slightly larger in SG stars. D'Antona et al. (2013) showed that understanding the CMD of the clusters in the dwarf galaxy Fornax requires the SG stars to lose more mass while ascending the RGB, and indicated a faster rotation rate as possible reason. Here, we suggest that fast rotation is responsible for the brighter luminosity of the LFM. Rotation will increase the mass evolving on the RGB, the total mass lost (Georgy et al. 2013) and the core mass necessary for the flash ignition (Mengel & Gross 1976). We run numerical simulations to test the hypothesis that the luminosity distribution of the hottest BHk stars originates from a spread in the core masses and, hence, of the rotation rates. To reproduce the observations, we assumed a Gaussian distribution of core masses; the centre of this distribution is increased with respect to the core mass of the standard, non-rotating models, in such a way that it corresponds to an $F225W$ flux one-half magnitude larger than the peak of the standard distribution. Because the cooler side of the EHB can easily be reproduced with small (if any) rotation, we must allow for a variation in the average speed of rotation in the transition from LFM to 'normal' HB models. Based on these arguments, we constructed a new synthetic population: we extracted a core mass according to a Gaussian distribution with $\Delta M_{\mathrm{core}} = 0.05$ M$_\odot$[8] and $\delta m_{\mathrm{core}} = 0.03$ M$_\odot$. Since different rotation rates among the clusters stars result in a dispersion in both core mass and total mass, we assumed that stars rotating below a limiting threshold are standard EHB, while those rotating faster are LFM. In this case, the limit was set at $\Delta M_{\mathrm{core}} = 0.01$ M$_\odot$, a value that, according Mengel & Gross (1976), corresponds to an average initial rotation of about $\omega = 0.000\,15$ s$^{-1}$. This choice of parameters allows us to reproduce the observations as shown in Fig. 6. We are aware that, before considering this as a plausible interpretation, it is necessary to consider in detail the uncertainties of the LFM models. This aspect will be addressed in a forthcoming work (Tailo et al., in preparation); what we want to emphasize here is that with our models, the magnitude extension of the BHk can be met only by assuming a larger helium core. The dispersion in angular frequency of the stars on the MS due to the extreme environment that occurs during the formation of the 'extreme' SG is an open issue on which our group is working.

We conclude this discussion with a comment on the sample of BHk stars spanning a range of ~2 mag below $m_{F225W} = 21.4$ that are not matched by the present simulations.

Fig. 7 shows the post-HB evolution of the least massive model not undergoing the late helium flash mixing. The tracks, before reproducing an AGB manque evolution, show a strong rise in brightness in conjunction with the transition from core to shell helium burning. This phase lasts approximately 20 Myr, and is sufficiently long to account for the aforementioned group of bright stars. Fig. 7 also shows (red line) the pre- and post-HB evolution of an LMF star. In the latter phase, the track moves directly to the WD cooling sequence, whereas during the pre-HB phase, the star evolves at magnitudes $19.5 < m_{F225W} < 21.5$. However, in this case, the duration of this phase is too short ($\sim 10^3$ yr) to have a meaningful statistical impact and it is very unlikely that the observed stars belong to this evolutionary phase.

---

[8] This choice of the parameters corresponds to an angular frequency of $\omega \sim 3.5 \times 10^{-4}$ s$^{-1}$ in MS stars according to Mengel & Gross (1976), which can easily be achieved (see for example Armitage, Clarke & Palla 2003).

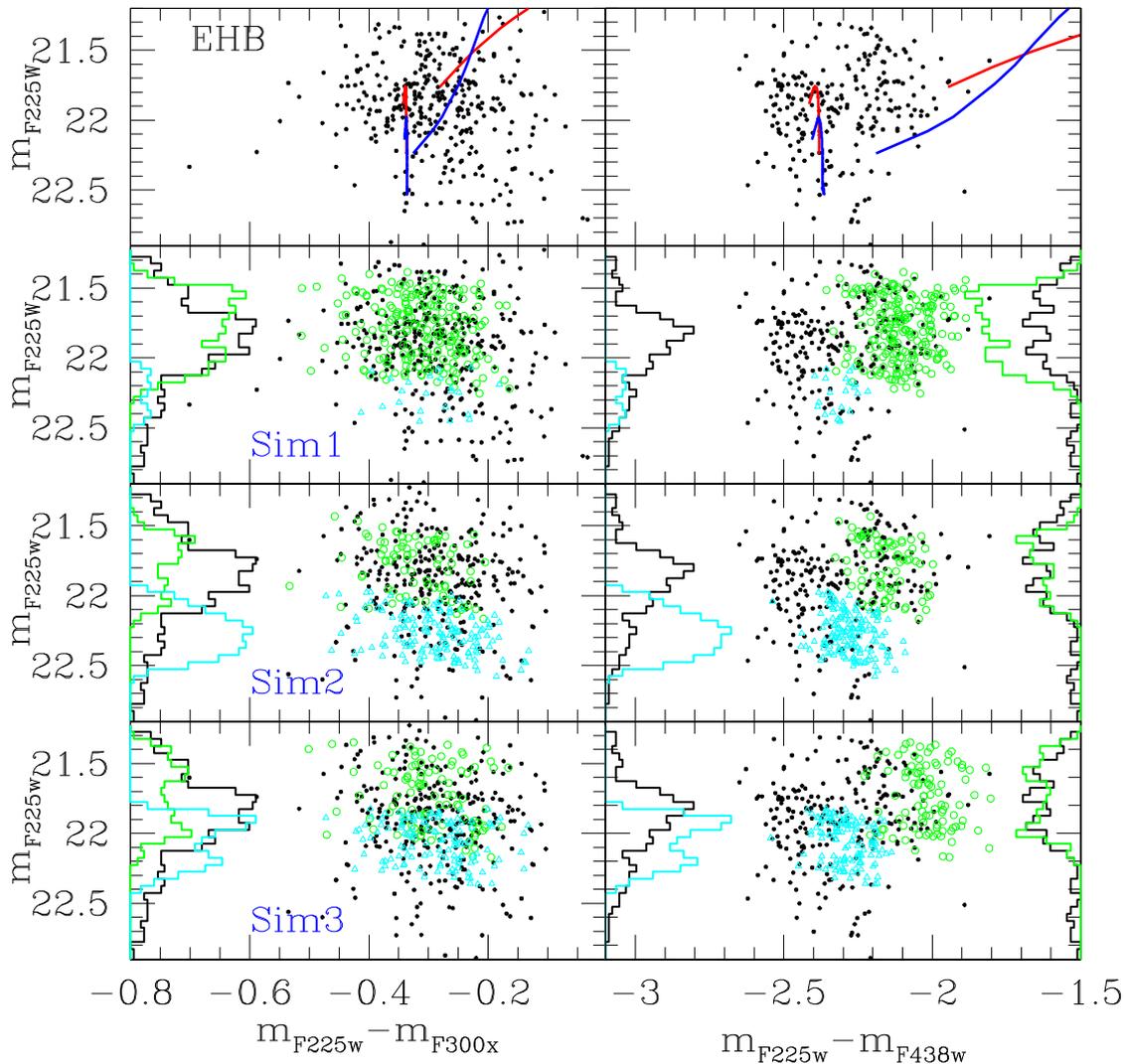

**Figure 5.** CMD of observed EHB stars (black closed circle) compared with the three different simulations described in the text and in Table 3 for $m_{F225W} - m_{F300X}$ (left-hand panels) and $m_{F225W} - m_{F438W}$ (right-hand panels) colours. The cyan triangle and open circle are LFM and 'normal' EHB, respectively. In the upper panels, where only observed stars are shown, the HB evolution of LFM stars and ZAHB with $Y = 0.25$ and 0.40 are displayed. The comparison between observations and simulations is done through the histograms in magnitude (black thick line for the observations and coloured thin for the simulations). For the CMDs in $m_{F225W} - m_{F438W}$ colours, the histograms of both LFM and EHB population are drawn.

## 5 CONCLUSIONS

The starting point of this work was to refine the analysis of the HB of NGC 2419 with the help of *HST* WFC3 UV photometry along with the indications given by Ventura et al. (2012) on the helium content of the extreme stars.

The observations allowed us to identify, with extreme clarity, three different populations in the HB of this cluster. In particular, using the $m_{F225W} - m_{F300X}$ colour, it is possible to recognize a brighter HB, further split into two subpopulations separated by a gap: a dominant part, with a strongly peaked distribution of stars with $Y = 0.25$, and a very small (9 per cent), brighter population, which we interpret as composed by stars with a very little spread in helium ($\Delta Y < 0.03$).

We have identified two distinct groups of stars along the EHB, which are clearly visible in the CMD when the $m_{F225W} - m_{F438W}$ colour is used: we find that this EHB is made up of helium-rich stars, half of which (the hottest part) have suffered a delayed helium-core flash on the WD cooling curve. Mixing between the convective surface and the helium burning core enriches the envelope of these stars in helium ($\sim 96$ per cent) and carbon ($\sim 2$ per cent by mass). This analysis allows us to determine the initial helium of the stars populating the EHB: we find $Y = 0.36$, in nice agreement with the predicted helium in the ejecta of massive AGBs of the same metallicity as NGC 2419 (Ventura et al. 2012). This result further supports the hypothesis that the SG stars in GCs formed from the ashes of the AGB evolution of intermediate-mass stars. However, there is a problem with this interpretation. As previously noted by Brown et al. (2010), the predicted brightness of LFM stars is about one-half magnitude brighter than that observed for BHk stars. To explain the poor agreement between the brightness of these stars and the luminosities of the 'LF mixed' models, we suggest that the core masses of LFM stars are larger than those expected on the basis of standard RGB evolutionary models. We speculate that a faster initial rotation rate among SG stars is the reason for this increase in the core mass. We can reproduce the observed distribution of HB stars in NGC 2419 by assuming that the progenitors of the non-LFM EHB stars rotate slowly whereas the bulk of progenitors of LFM stars, owing to a high rotation rate, evolve with core masses on the average ∼0.05 M$_\odot$ larger. Following the computations by Mengel & Gross (1976), we suggest an average initial rotation of about $\omega \sim 3.5 \times 10^{-4}$ s$^{-1}$, which can easily be achieved on the MS. If further data confirm this interpretation, a fast initial rotation would be a strong signature of the peculiarity of extreme SG stars in GCs.

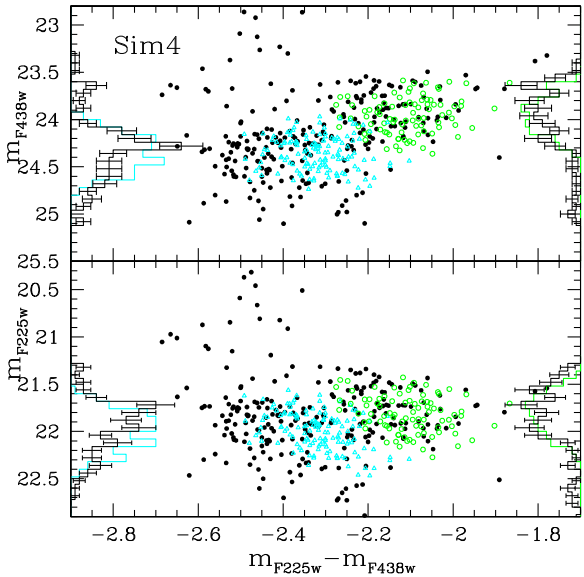

**Figure 6.** As Fig. 5 but for Sim4 where a Gaussian distribution for both core mass ($\Delta M_{core} = 0.05$) and total mass ($\Delta M = 0.195$) is assumed in order to simulate a dispersion in angular velocities of SG stars. The further assumption is that the only stars that rotate slowly are the EHB stars. As in Fig. 5, the quantitative comparison is shown in the observed (black) versus simulated (cyan and green) histograms of counts as a function of magnitude. The error bars on the observed histogram are the result of individual count Poisson errors $\sqrt{N}$.

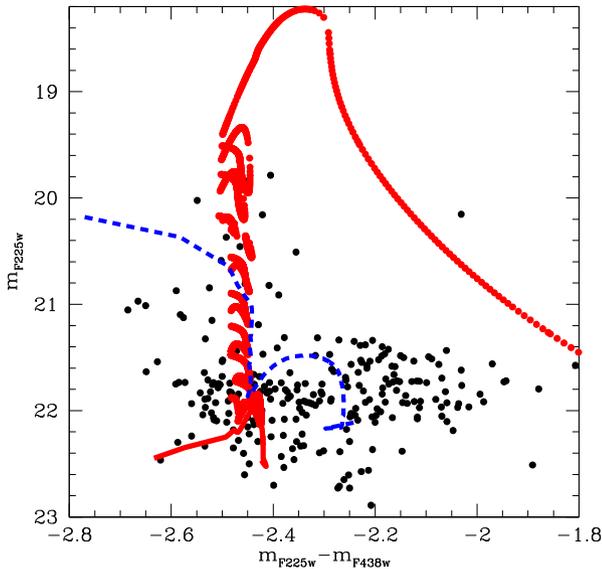

**Figure 7.** Possible interpretations of very luminous BHk stars (see the text). Dashed (blue) line shows the post-HB evolution of the M = 0.496 M$_\odot$ model while the thick (red) line represents the pre- and post-HB evolution of an LFM star.

## ACKNOWLEDGEMENTS

We thank the referee for providing constructive comments and T. Brown to for having made available his spectra.

## REFERENCES

Adams T., Davies M. B., Sills A., 2004, MNRAS, 348, 469
Anderson J., Bedin L. R., Piotto G., Yadav R. S., Bellini A., 2006, A&A, 454, 1029
Armitage P. J., Clarke C. J., Palla F., 2003, MNRAS, 342, 1139
Beccari G., Bellazzini M., Lardo C., Bragaglia A., Carretta E., Dalessandro E., Mucciarelli A., Pancino E., 2013, MNRAS, 431, 1995
Bedin L. R., Cassisi S., Castelli F., Piotto G., Anderson J., Salaris M., Momany Y., Pietrinferni A., 2005, MNRAS, 357, 1038
Bekki K., 2012, ApJ, 747, 78
Bellini A., Bedin L. R., Piotto G., Milone A. P., Marino A. F., Villanova S., 2010, AJ, 140, 631
Bellini A., Anderson J., Bedin L. R., 2011, PASP, 123, 622
Brown T. M., Sweigart A. V., Lanz T., Landsman W. B., Hubeny I., 2001, ApJ, 562, 368
Brown T. M., Sweigart A. V., Lanz T., Smith Ed L., Wayne B., Hubeny I., 2010, ApJ, 718, 1332
Cardelli J. A., Clayton G. C., Mathis J. S., 1989, ApJ, 345, 245
Carretta E., Gratton R. G., Bragaglia A., D'Orazi V., Lucatello S., Sollima A., Sneden C., 2013, ApJ, 769, 40
Cassisi S., Salaris M., Anderson J., Piotto G., Pietrinferni A., Milone A., Bellini A., Bedin L. R., 2009, ApJ, 702, 1530
Castellani M., Castellani V., 1993, ApJ, 407, 649
Chung C., Yoon S.-J., Lee Y.-W., 2011, ApJ, 740, L45
Cohen J. G., Kirby E. N., 2012, ApJ, 760, 86
D'Antona F., Caloi V., 2004, ApJ, 611, 871
D'Antona F., Caloi V., 2008, MNRAS, 390, 693
D'Antona F., Caloi V., Ventura P., 2010, MNRAS, 405, 2295
D'Antona F., Caloi V., Montalbán J., Ventura P., Gratton R., 2002, A&A, 395, 69
D'Antona F., Caloi V., D'Ercole A., Tailo M., Vesperini E., Ventura P., Di Criscienzo M., 2013, MNRAS, 434, 1138
D'Cruz N. L., Dorman B., Rood R. T., O'Connell R. W., 1996, ApJ, 466, 359
D'Cruz N. L. et al., 2000, ApJ, 530, 352
D'Ercole A., Vesperini E., D'Antona F., McMillan S. L. W., Recchi S., 2008, MNRAS, 391, 825
di Criscienzo M. et al., 2011a, AJ, 141, 81 (DC11)
di Criscienzo M. et al., 2011b, MNRAS, 414, 3381 (DC11)
Dorman B., O'Connell R. W., Rood R. T., 1993, BAAS, 25, #101.03
Dorman B., O'Connell R. W., Rood R. T., 1995, ApJ, 442, 105
Georgy C. et al., 2013, A&A, 558, 103
Greggio L., Renzini A., 1990, ApJ, 364, 35
Heber U., Kudritzki R. P., Caloi V., Castellani V., Danziger J., 1986, A&A, 162, 171
Holtzman J. A., Burrows C. J., Casertano S., Hester J. J., Trauger J. T., Watson A. M., Worthey G., 1995, PASP, 107, 1065

Ibata R., Sollima A., Nipoti C., Bellazzini M., Chapman S. C., Dalessandro E., 2011, ApJ, 743, 43
Lanz T., Brown T., Sweigart A., Hubeny I., Landsman W., 2004, ApJ, 602, 342
Mengel J. G., Gross P. G., 1976, Ap&SS, 41, 407
Miller-Bertolami M. M., Althaus L. G., Unglaub K., Weiss A., 2008, A&A, 491, 253
Milone A. P., Stetson P. B., Piotto G., Bedin L. R., Anderson J., Cassisi S., Salaris M., 2009, A&A, 503, 755
Milone A. P. et al., 2012, A&A, 540, 16
Moni Bidin C., Moehler S., Piotto G., Momany Y., Recio-Blanco A., 2009, A&A, 498, 737 (M80)
Mucciarelli A., Bellazzini M., Ibata R., Merle T., Chapman S. C., D'Alessandro E., Sollima A., 2012, MNRAS, 426, 2889
Pasquato M., de Luca A., Raimondo G., Carini R., Moraghan A., Chung C., Brocato E., Lee Y.-W., 2014, ApJ, 789, 28
Ripepi V. et al. 2007, ApJ, 667, L61
Salaris M., Cassisi S., Pietrinferni A., 2008, ApJ, 678, 23
Shetrone M. D., Cote P., Sargent W. L. W., 2001, ApJ, 548, 592
Ventura P., D'Antona F., Mazzitelli I., Gratton R., 2001, ApJ, 550, L65
Ventura P., Caloi V., D'Antona F., Ferguson J., Milone A., Piotto G. P., 2009, MNRAS, 399, 934
Ventura P., D'Antona F., Di Criscienzo M., Carini R., D'Ercole A., Vesperini E., 2012, ApJ, 761, L30
Ventura P., Dell'Agli F., Schneider R., Di Criscienzo M., Rossi C., La Franca F., Gallerani S., Valiante R., 2014, MNRAS, 439, 977

This paper has been typeset from a T$_{\rm E}$X/L$^{\rm A}$T$_{\rm E}$X file prepared by the author.